\documentclass{ws-mpla}

\begin{document}

\markboth{C. M. G. Sousa}{On the frequency shift of gravitational waves}

\catchline{}{}{}{}{}

\title{ON THE FREQUENCY SHIFT OF GRAVITATIONAL WAVES}

\author{\footnotesize CLAUDIO M. G. DE SOUSA}

\address{Universidade Federal do Para, Faculdade de Fisica Ambiental, \\
               Av. Marechal Rondon, s/n, Santarem, PA 68040-070, State of Para,
                Amazon region, Brazil.\footnote{Main address}
              \\ ( claudiogomes@ufpa.br , claudio@unb.br )
              \\ {\em and} \\
              Universidade Catolica de Brasilia, Diretoria de Fisica e Matematica, \\ QS 07 Lt 01 EPCT Aguas Claras,
              Brasilia, DF 72030-170, Brazil.
}

\maketitle

\pub{Received (Day Month Year)}{Revised (Day Month Year)}

\begin{abstract}
Considering plane gravitational waves propagating through flat spacetime,
it is shown that curvatures experienced both in the starting point and 
during their arrival at the earth can cause a considerable shift in
the frequencies as measured by earth and space-based detectors.
Particularly for the case of resonant bar detectors 
this shift can cause noise-filters to smother the signal.
\end{abstract}

\ccode{PACS Nos.: 04.30Nk, 95.55.Ym.}

\section{Introduction}	

The evidence that the decay of a binary pulsar \cite{Taylor78} is related to
the emission of gravitational waves shows that the predictions of general
relativity are valid even if we use approximations as those of the
linearized models for small perturbations in the metric tensor.

Despite the claim that gravitational waves have never been directly
detected, the study of this subject has afforded large development both in
theoretical and experimental fields, not only in general relativity, but
also in detectors technology, noise treatment, etc.

Nevertheless, direct detection of gravitational waves can unfold several
mysteries of modern cosmology. As an example, the detection of very-long
wavelength gravitational waves can improve the knowledge about the origin of
the cosmic microwave background \cite{Caldwell99}. Another example is that
of gravitational waves from neutron-star--black-hole spiral that could
bound the strength of a scalar field in scalar-tensor theories of gravity 
\cite{Will02}, since in those theories there exists the possibility of a
dipole gravitational radiation. Astrophysical characteristics can also be
inferred from the measurements of gravitational waveforms \cite{Vallisneri00,Saijo00}. 
It is desirable, thence, that we have a precise idea of
gravitational radiation spectra for different sources.

This paper intents to verify the consequences of a frequency shift on
gravitational waves that could cause detectors to be working out of their
optimum. It is argued here that with few corrections modern detectors will
increase their probability of detecting directly gravitational waves.

The foundation of this shift effect is in the Einstein Equivalence Principle \cite{Will93} (EEP): for instance,
if a gravitational wave approaches the earth with a frequency $f$ it could
only be detected with the same frequency by a free falling antenna. The
inverse problem occurs with gravitational waves outgoing from the source.
The influence upon gravitational radiation by its own source is known in the backscattering
theory, which involves detailed knowledge of the quadrupole moment of the
source \cite{Isaacson68,Thorne80}.
The fact that gravitational waves propagation is similar to electromagnetic 
waves has been reported by several authors in the backscattering arena 
\cite{Campbell73,Malec01}. It is also known the fact that background curvature
affects the propagation of the gravitational waves from stochastic cosmological sources \cite{Durrer06}.
Despite backscattering theory has increased with the discussion of a possible suppression 
of gravitational waves \cite{Price92,Kundu93}, and the subsequent demonstration 
of the absence of such $\omega M$ effects, it is not on the purpose of the present 
work to show any suppression on gravitational waves, but only the consequences of its 
frequency deviation to earth based resonant detectors.

In this paper it is also supposed that to quantify the frequency shift effect and
to determine if it is worthwhile to be considered, it is not required the
full nonlinear theory, and one can extend the geometric optics to regions
where the background influences strongly the gravitational radiation.
Considering wavelengths short compared to the radius of curvature of the
background space-time, one can observe that the propagation equations for
the gravitational radiation yields similar phenomena as those observed for
the electromagnetic radiation \cite{Misner73}.

Using the metric: 
\begin{equation}
g_{\mu\nu}=\eta_{\mu\nu}+h_{\mu\nu},  \label{E1}
\end{equation}
where $h_{\mu\nu}$ represents a small perturbation on the flat spacetime
metric $\eta_{\mu\nu}$ (with $\| h_{\mu\nu} \| \ll 1$), the linearized
vacuum field equations take the form: 
\[ 
\Box h_{\mu\nu} + h,_{\mu\nu}- h^\alpha_{\nu ,\mu\alpha} - h^\alpha_{\mu
,\nu\alpha} =0. 
\]

But, with the choice of the gauge \cite{Schutz01}: 
\[
h^\alpha_{\mu ,\alpha} - \frac{1}{2} h,_\mu =0, 
\]
where $h=\eta^{\mu\nu}h_{\mu\nu}$, one obtains: 
\begin{equation}
\Box h_{\mu\nu} =0.  \label{E3}
\end{equation}

As first attempt to compute the effects of the gravitational field on the
gravitational wave in the very moment it comes out from the wave generation zone
close to the source, the plane wave formalism will be extended even to
regions where the field is strong.

Hence, consider the gravitational radiation propagating as a plane wave
given by some solution of the equation~(\ref{E3}), in an eikonal type form: 
\begin{equation}
h_{\mu\nu} = Re\left\{ A_{\mu\nu} e^{ikx}\right\},  \label{E4}
\end{equation}
where $k \cdot x = k_0 x^0 + k_i x^i = \Psi$ (with $i=1,2,3$) and $\Psi$ is
the eikonal from which the wave vector ${\bf {\rm k}} = -{\bf \nabla} \Psi$
and the frequency $\omega =\Psi,_0$ (i.e., $k_\mu \equiv \Psi,\mu$) can be
obtained.

Considering $A^{*}_{\mu\nu}$ as the complex conjugate of $A_{\mu\nu}$, one
can also define the scalar amplitude 
$A=\sqrt{\frac{1}{2}A^{*}_{\mu\nu}A_{\mu\nu}}$ 
and the polarization ${\displaystyle e_{\mu\nu}=\frac{A_{\mu\nu}}{A}}$.

Using the above defined quantities one can develop all geometrical optics
formalism for the gravitational waves \cite{Sabbata77}. The analogy between
gravitational waves and light using the eikonal (WKB\ approximation) was
 addressed by Isaacson \cite{Isaacson68}. This approximation becomes
more accurate for high-frequencies, when the wavelength is small compared to
the radius of curvature of the background geometry, and the Riemann and the
Ricci tensors expansions are gauge invariants to an extremely good
approximation, since geometry can be considered as locally flat over
distances of order $L$, where in first approximation one gets $R_{\mu \nu
\alpha \beta }^{(0)}\sim L^{-2}$. Thus considering $\lambda \ll L$ one ensures
that $A_{\mu \nu }$ and $k_{\mu }$ vary slowly over a characteristic
distance of order $L$.

But the main feature to be considered here is the propagation frequency of
gravitational waves in such way that, in the next section, we can analyze
the effects of the source on its own gravitational wave frequency.

Noticing that the proper-time is \cite{Einstein,Landau}: 
\begin{equation}
t= \int \sqrt{g_{00}} dx^0,  \label{E5}
\end{equation}
we have: 
\[
\frac{\partial}{\partial t}h_{\mu\nu} = \frac{\partial h_{\mu\nu}}{\partial
x^0}\frac{\partial x^0}{\partial t} = h_{\mu\nu ,0}\frac{1}{\sqrt{g_{00}}} , 
\]
and so: 
\begin{equation}
\omega = \frac{\omega_0}{\sqrt{g_{00}}} .  \label{E7}
\end{equation}
Since $g_{00} = 1+2\phi /c^2$ (recovering natural units, where $c$ is the
velocity of light), where $\phi <0$ is the gravitational potential (so
normalized that $\phi (\infty )=0$), in the linearized theory one can write
the approximation: 
\[
\omega = \omega_0 \left( 1-\frac{\phi}{c^2} \right) . 
\]

\begin{figure}[h]
\centerline{\psfig{file=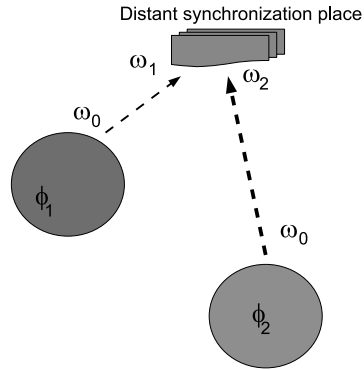, width=2.0in, angle=0}}
\vspace*{8pt}
\caption{Scheme used to obtain equation (\ref{E9}) showing the frequency synchronization of two different gravitational potentials, $\phi_1$ and $\phi_2$. Similar experiments are carried out on the planets emitting signals with the same frequency $\omega_0$, which are influenced by the gravitational potentials. This experiment allows one to calibrate $\Delta\omega$.
 \protect\label{fig3}}
\end{figure}

If the gravitational radiation is emitted (Fig.\ref{fig3}) from a point where the potential
is $\phi _{1}$ with frequency $\omega $, and is received at a point where
the potential is $\phi _{2}$, then the frequency is shifted from its
original value in $\Delta \omega $, given by: 
\begin{equation}
\Delta \omega =\left( \frac{\phi _{1}-\phi _{2}}{c^{2}}\right) \omega ,
\label{E9}
\end{equation}
which is the first-order Doppler effect formula usually found in
electromagnetic redshift investigation. 
Hence, gravitational waves experience a Doppler-like effect due to the
difference between the source and the earth gravitational potentials
(neglecting relative velocities).

For instance, taking $\phi _{1}$ as the gravitational potential of the
source and $\phi _{2}$ on the earth, most of the detectable events will
happen with $\mid \phi _{1}\mid \gg \mid \phi _{2}\mid $ and, hence, $\Delta
\omega <0$, showing that there is an overall redshift on gravitational waves
to be detected on the earth (except for background gravitational waves that
baths the space). Thus, during its travel from the source to a ground based
detector gravitational waves will usually be found in a lower frequency than that
expected by standard calculations.

\section{Sample Calculations}

Some of the expected events in the search of gravitational radiation can be
selected to determine if corrections are demanded for modern detectors.

For the purpose of a simple analysis one can use equation (\ref{E9}) to
write: 
\begin{equation}  \label{E10}
\frac{\Delta f}{f}=\frac{\phi_1 - \phi_2}{c^2}
\end{equation}
where $f$ is the source frequency, which has potential $\phi_1$.

Typical earth mass and radius give rise to $\phi_2 \simeq -6\times 10^7$%
J/kg, negligible compared to the usual gravitational potential for the
target sources.

As two examples, consider the cases of a collapsing neutron star and of a
black hole. If one considers a collapsing neutron star with $M_{{\rm NS}}\simeq M_\odot$
and $R_{{\rm NS}}\simeq 10^4$m, and so $\phi_1 =\phi_{{\rm NS}}\simeq
-1.3\times 10^{16}$J/kg, the range \cite{Muller82}
of characteristic frequencies is $10^3\leq f \leq 10^4$Hz.
Another example to be considered is that of black holes, which have mass 
$2M_\odot \leq M_{{\rm BH}}\leq 10^{10}M_\odot$ and, considering the horizon 
$R_{{\rm BH}}\simeq 2M_{{\rm BH}}$, the exact mass of the black
hole does not affect the final result for the potential which is always 
$\phi_{{\rm BH}}\approx -G/2$.

Now, computing the radiation shift for collapsing neutron star case with the use of
equation (\ref{E10}), one obtain: 
\[
\Delta f_{{\rm NS}}\simeq -\frac{\phi_{{\rm NS}}}{c^2}f_0 
\]
where $f_0$ is the original source frequency. Hence, the gravitational
radiation for the collapsing neutron star suggested above gives: 
\begin{equation}  \label{E12}
\frac{f_{{\rm NS}}}{f_0}\simeq 0.8517
\end{equation}
and the frequency range is no longer $10^3\leq f \leq 10^4$Hz, but $852\leq
f \leq 8517$Hz.

As can be expected, the effect is stronger the higher is the gravitational
potential of the source. For black holes $\Delta f \simeq -\frac{1}{2}f_0$,
or: 
\begin{equation}  \label{E13}
f_{{\rm BH}}=\frac{f_0}{2}
\end{equation}

Some known results \cite{Hawking87} are presented on Table~\ref{Table1} 
with their redshift values predicted by this effect, considering
the shifts are caused only by geometries of the source and of the earth. For the
compact binary on the Table~\ref{Table1} the mass is approximately $M_{\odot }$ 
and radius $R_{\odot }$, and the expected redshift is negligible, approximately $10^{-6}$. 
The stochastic gravitational wave background presents a negligible
blueshift for earth-based detectors. Furthermore, the characteristic
frequencies for the stochastic GW background ($10^{-5}$ to $0.03$ Hz) is in
the low frequency regime, where the approximations considered here fail. The
redshifts have been computed using $1+z=f_{em}/f_{ob}$, with emitted and
observed frequencies, $f_{em}$ and $f_{ob}$, respectively.

\begin{table}[h]
\tbl{Approximate frequency of some sample sources usually considered as
targets for gravitational waves search and the expected shifts caused by the
gravitational field of the source and of the earth. The gravitational
potential for the black hole is in geometrized units. Note that for the
example given for coalescence of compact binaries the redshift is
negligible. For the stochastic background there is a negligible blueshift.}
{ \begin{tabular}{@{}|l|c|c|c|c|c|@{}} \hline
Sample & Source Grav. & Emitted & Expected & Shift & z \\ 
Description & Potent.(J/kg) & Freq. (Hz) & Freq. (Hz) & (Hz) &  \\ 
\hline\hline
Collapsing &  &  &  &  &  \\ 
Neutron Star & $-1.3\times 10^{16}$ & $10^4$ & $8.5 \times 10^3$ & 14.8\% & 
0.1741 \\ \hline
Collapsing &  &  &  &  &  \\ 
Black Hole & $-G/2 $ & $10^6$ & $5 \times 10^5$ & 50.0\% & 1 \\ \hline
Coalescence &  &  &  &  &  \\ 
of Compact & $-5.34\times 10^{11}$ & $10^2$ & $\simeq 10^2$ & $\simeq$ 0.0 & 
$\simeq$ 0 \\ 
Binaries &  &  &  &  &  \\ \hline
Stochastic &  &  & nearly &  &  \\ 
Background & -- & $10^{-5}$ -- $0.03$ & unchanged & $\simeq$ 0.0 & $\simeq$ 0 \\
\hline
\end{tabular} \label{Table1} }
\end{table}

\section{Resonant detectors}

Gravitational waves can experience almost all the characteristics of
electromagnetic waves, apart from the properties related to the charges that
originated the wave. Together with those characteristics is the shift of
gravitational wave frequency, which is a Doppler-like effect for
gravitational waves. But if gravitational waves have their frequency
shifted, what could be the consequences for detectors?

Focusing on resonant detectors, which are known as narrow-band around the
frequency $f_0$ they are looking for, the signal-to-noise ratio is 
given by \cite{Hawking87}: 
\begin{equation}  \label{E14}
\frac{S^2}{N^2}=\frac{\frac{\pi}{2}f_0^2 \mid \tilde{h}(f_0)\mid^2}{kT_n}%
\int\sigma_0 (f)df \, ,
\end{equation}
where $\tilde{h}(f_0)$ is the Fourier transform of $h(t)$, $T_n$ is the
noise temperature (which takes care of all possible noises), and $\sigma_0$
is the cross section of the detector. Typically the noise is so severe that
it becomes necessary to use a bandwidth $\Delta f$ very small compared to
the sought frequency $f_0$. Hence, if one considers the case of the black
holes in equation (\ref{E14}), the original signal-to-noise ratio 
$(SNR)_0$ can roughly reach: 
\begin{equation}  \label{E15}
SNR \simeq \frac{1}{4} (SNR)_0 \, .
\end{equation}
Neglecting further effects of the 50\% shift on $\tilde{h}(f_0)$, the noise
smothers source signal. 
To better understand the device behaviour with a biased signal,
consider resonant detectors  as EXPLORER and NAUTILUS.
With $SNR=1$, the gravitational wave spectrum is given by \cite{Pizzella01}:
\begin{equation} \label{E14a}
S_h(\omega )= \pi^2 \frac{k T_e}{M Q v^2} \frac{\omega_0^3}{\omega^4}   %
\left( 1 + \Gamma  \left(  Q^2  \left( 1 - \frac{\omega^2}{\omega_0^2}  %
\right)^2  \frac{\omega^2}{\omega_0^2}   \right)   \right)
\end{equation}
where $k$ is the Boltzmann constant, $T_e$ is the equivalent temperature 
which includes backreaction for the 
electronic amplifier, $M$ is the cylinder mass, $Q$ is the quality factor of the
bar material, $v$ is the sound velocity in the bar material, $\Gamma$ is a characteristic 
dimensionless factor (usually very small), $\omega$ is the incident frequency 
and $\omega_0$ is the bar natural resonance frequency.

For $\omega =\omega_0$ the detector has its highest sensitivity:
\begin{equation} \label{E14b}
S_h(\omega_0 )= \pi^2 \frac{k T_e}{M Q v^2} \frac{\omega_0^3}{\omega^4}   %
\left( \frac{1 + \Gamma}{\omega_0}   \right) 
\end{equation}

The spectral amplitude is given by:
\begin{equation} \label{E14c}
\tilde{h}=\sqrt{S_h}
\end{equation}

For EXPLORER and NAUTILUS one can use $T_e=0.1$K, $M=2270$kg, $Q=8.5\times 10^5$, $v=5400$m/s (sound velocity in aluminum) and $\Gamma = 10^{-6}$.

\begin{figure}[h]
\centerline{\psfig{file=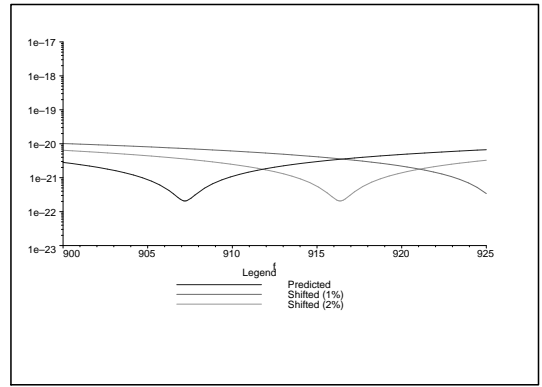,width=2.0in,angle=270}}
\vspace*{8pt}
\caption{Spectral amplitudes $\tilde{h}$ versus source frequency. 
Thick line represents the predicted curve using $f=f_0$, with minimum at $907.2$Hz. 
The minimum of the curve corresponds to maximum sensitivity, and is of order $10^{-22}$,
as expected for Explorer and Nautilus.
When very small shifts are introduced the minimum is displaced (thin lines).
  .\protect\label{fig1}}
\end{figure}

Figure~\ref{fig1} shows plots for spectral amplitudes $\tilde{h}$ using eq.(\ref{E14a}). The thick line corresponds to the predicted spectrum when the wave reaches detector with the same frequency $\omega$ as that of the bar resonance $\omega_0$ (Figure~\ref{fig1} used $f=f_0=907.2$Hz). The minimum corresponds to expected point of maximum sensitivity of the device, and is of order $10^{-22}$Hz$^{-1/2}$. Notice that for slightly shifted values of gravitational waves frequency major modifications are observed on the point of maximum sensitivity.

\begin{figure}[h]
\centerline{\psfig{file=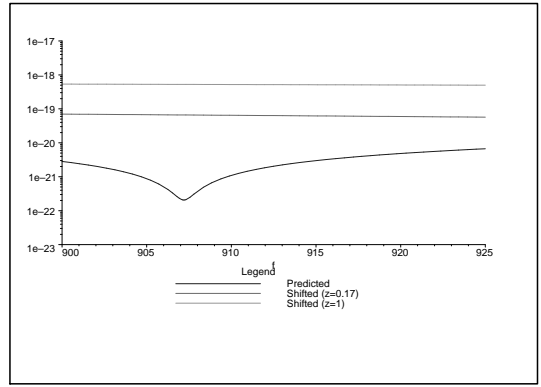, width=2.0in, angle=270}}
\vspace*{8pt}
\caption{Spectral amplitudes $\tilde{h}$ versus source frequency for the samples collapsing 
neutron star ($z=0.17$) and collapsing black hole ($z=1$). 
Thick line represents the predicted curve using $f=f_0$. 
The upper line corresponds to case $z=1$ and the middle line corresponds to case $z=0.17$.
In Explorer and Nautilus operating frequency range, there is a huge change in the sensitivity
 for some of the most expected events.
 \protect\label{fig2}}
\end{figure}

Figure~\ref{fig2} represents the same scenario for the samples collapsing neutron star ($z=0.17$) and collapsing black hole ($z=1$). The sensitivity is of order $10^{-20}$Hz$^{-1/2}$ and $10^{-19}$Hz$^{-1/2}$, respectively.
Nevertheless, this does not mean that detectors are unable to receive signals from sources with different frequencies. It is just matter of time receiving the signal that correctly matches with the sensitivity. 
However, such frequency shift can seriously decreases the probability of detecting 
most of the high-intensity sources in the range devices has been projected to.

Similar calculations can be carried out to verify
what are the consequences of this effect for broad-band detectors, as
GEO600, VIRGO, LIGO and TAMA, which operate as laser interferometric
detectors at room temperature, and thermal and shot noise become larger
compared to the signal \cite{Virgilio01}, since the power output is proportional to $f^{1/2}$.

Besides the highly
sophisticated noise treatment \cite{Schutz97} necessary to analyze GW
detector output, one must extract the filtered signal after considering many
factors. For instance, orbital motions of the LISA detector will cause the
appearance of a frequency modulation \cite{Cornish02}, which can be used to
determine the source location and orientation.
LISA will work in the low-frequency band, $10^{-4}$--$10^{-1}$Hz (which, due
to seismic noise, is not reached by the ground-based detectors). On the
other hand, in LIGO detectors \cite{Fritschel01} the filtering of seismic
noise is so heavy that it creates a cutoff frequency around 10Hz at the test
masses, in such that some events predicted to fall nearby that cutoff
frequency can be missed.

\section{Limits of Validity}

The fact that EEP has been mentioned does not impose the use of a reference frame
with no curvature. An event detected by an antenna is considered 
as an experiment. Due to EEP, an experiment in a free falling laboratory will give same result 
as if it were in empty space. The fact that the spacetime is locally flat does not contradict with
the possibility of small perturbations to propagate, if ripples
influence on the background curvature is considered negligible (which is usual for
high frequency waves)\cite{Misner73}. 

Geometric optics \cite{Mashhoon86} requires a large $\Psi$ and that 
$\raisebox{.6ex}{-}\!\!\!{\lambda} \ll \ell (x)$, 
which is to say the (reduced) wavelength must be very small compared to the radius of 
curvature of local background. 
Rewriting (\ref{E1}) in the form $g_{\mu\nu}=g_{\mu\nu}^{\rm B} + h_{\mu\nu}$. 
The propagation \cite{Misner73} is due to 
$h_{\mu\nu}$ and after skipping the generating zone, it is considered that the waves reach empty
space. For high-frequencies the influence of the waves on $g_{\mu\nu}^{\rm B}$ is considered 
negligible.
Hence, expressions (\ref{E7})-(\ref{E9}) consider 
the background influence on gravitational waves only in the generating zone, and use 
flat background for the propagation zone.

Discussing further the analysis carried out by Mashhoon\cite{Mashhoon86} concerning the propagation of electromagnetic waves using geometrical optics on a  curved manifold, some remarks become necessary in such to ensure the validity of the present geometric optics approximation analogy for gravitational waves: 

(i) Two observers moving apart each other induce a frequency Doppler shift 
\[\raisebox{.6ex}{-}\!\!\!{\lambda}'(x) = \sqrt{\frac{1+\beta}{1-\beta}} \,
\raisebox{.6ex}{-}\!\!\!{\lambda}(x)
\]
where $\beta \geq 0$ is the $x$-axis velocity of the observer ${\cal O}'$ with respect to a stationary observer ${\cal O}$ at the origin ({\em i.e.}, $\beta =v/c$, and $c=1$ in geometrized units). 
Planets in Fig.\ref{fig3} are imagined to have $\beta \rightarrow 0$, in such that the standard Doppler effect can be neglected;

(ii) Mashhoon's work shows the validity limits of the $\raisebox{.6ex}{-}\!\!\!{\lambda} \ll \ell (x)$
assumption in two key cases: exterior Schwarzschild geometry, and the Friedmann-Lama\^{i}tre-Robertson-Walker model universe. A possible conflict appears if $\beta\rightarrow 1$, which can be circumvented using the geometric optics limit, and is here ensured by considering $\beta\rightarrow 0$. 

(iii) Finally, one must consider the restrictions on the frequency that can be imposed by the quantum limit for the propagation. Since the source gravitational potential is of order $\phi \sim G\hbar\omega / \raisebox{.6ex}{-}\!\!\!{\lambda} \sim (L_P / \raisebox{.6ex}{-}\!\!\!{\lambda})^2$, where $L_P$ is the Plank length, the influence of the background on the field can be neglected if $\phi \ll 1$ (in geometric units). To ensure this quantum limit does not affect the background for the propagation of the gravitational waves, it is necessary to consider wavelengths such that $\raisebox{.6ex}{-}\!\!\!{\lambda} \gg L_P$, which corresponds to an upper limit for the frequency, $f\ll 10^{43}$Hz. Nevertheless, if $\raisebox{.6ex}{-}\!\!\!{\lambda} \sim L_P$, quantum effects become considerable.

Despite Mashhoon have evaluated calculations for electromagnetic waves, there is a natural extension for gravitational waves preserving the covariant formalism, and the same limits apply.

\section{Conclusion}

Using the so-called geometrical optics for gravitational waves 
it is argued that
the difference between the gravitational potentials of
the source and the earth can cause shifts on gravitational radiation
frequencies, which are sometimes considerable. To quantify the
consequences of this effect geometric optics formalism has been extended to
regions where the background curvature is strong. 
 In the first order approximation carried
out here, one can see that very strong sources as black holes can have their
frequency shifted to 50\% of the original expected. For some kinds of
detectors this shift can cause noise-filters to smother the signal.

The resulting shift on gravitational waves frequency does not reduce the
importance of the presently existing devices.
Moreover, it is possible to increase their probability to detect directly
gravitational waves bursts if an adequate settlement of experiments take
into account the displacement of the original frequency expected. It just
emerges from this that another Doppler modulation factor should be taken
into account during detector pattern analysis.

Recently, it has been reported \cite{Rocchi04} that two cryogenic resonant 
detectors \cite{Astone01}  displayed coincident burst events 
in 1998 and in 2001: EXPLORER operates at frequencies ranging from 
904.7~Hz to 921.3~Hz and at a thermodynamic temperature of 2.6K; NAUTILUS 
operates at 906.97~Hz to 922.46~Hz and at a temperature of 1.5K. In 2001,
these detectors have obtained six coincidences in five days, approximately 
one coincidence per day, with an event signal corresponding to a burst with 
amplitude $h\simeq 2 \times 10^{-18}$. The correlation coefficient is 0.96,
detectors are 600km apart 
and, the connection with present paper is that during those runnings
 {\em no energy filter was applied}. 
In the reports concerning data obtained from EXPLORER and NAUTILUS,
the group preferred to take a
conservative position and testing the possibility of the events being 
result of cosmic ray interference or the possibility of accidental 
coincidences. The events are reported to be located in the direction of the 
center of the Galaxy, the same place Weber \cite{Weber70}
claimed to be the origin of signals detected in the seventies. 
The correct choice of the filters (and even the lack of them) 
could have improved EXPLORER and NAUTILUS detection 
sensitivity to adjacent frequencies.

\section{Acknowledgments}

I would like to thank L. Ph. Vasconcelos, H. Nazareno, A. Adib and R. Howell for 
reading the manuscript, and Department of Physics and Astronomy,
Dartmouth College, for the hospitality during the initial development of this work. 
I have also greatly benefited from stimulating discussions with M. D. Maia.
This work has been supported by CNPq (Conselho Nacional de Desenvolvimento 
Cientifico e Tecnologico), Brazil, under the contract 201031/96-5.

\end{document}